\documentclass[prb,aps,twocolumn]{revtex4}
\usepackage{amsmath}
\usepackage{amssymb}
\usepackage{bm}
\usepackage{epsfig}
\usepackage{graphicx}

\begin{document}

\title{Magnetic field effects on spin relaxation in heterostructures}

\author{M.M.~Glazov}\email{glazov@coherent.ioffe.ru}
\affiliation{A. F. Ioffe Physico-Technical Institute, Russian
Academy of Sciences, 194021 St. Petersburg, Russia}

\begin{abstract}
Effect of magnetic field on electron spin relaxation in quantum
wells is studied theoretically. We have shown that Larmor effect
and cyclotron motion of carriers can either jointly suppress
D'yakonov-Perel' spin relaxation or compensate each other. The
spin relaxation rates tensor is derived for any given direction of
the external field and arbitrary ratio of bulk and structural
contributions to spin splitting. Our results are applied to the
experiments on electron spin resonance in SiGe heterostructures,
and enable us to extract spin splitting value for such quantum
wells.
\end{abstract}

\maketitle

\section{Introduction}
Spin dynamics of charge carriers in semiconductors and
semiconductor nanostructures attract much attention during the
last decade. Theoretical and experimental efforts are made in
field of creation, conservation and manipulation of spin polarized
electrons. In this regard the control of spin dephasing processes
is a task of prime importance.

 The main spin relaxation mechanism
in high quality semiconductor quantum wells is that proposed by
D'yakonov and Perel' \cite{ivch_book,agw_rev}: electrons lose
their initial spin during free motion between scattering events
due to wavevector dependent effective magnetic field.\cite{dp,dk}
This field is determined by the conduction band spin splitting in
the systems lacking an inversion center. In bulk zinc-blende
lattice semiconductors the spin splitting is cubic in electron
wavevector $\bm k$ (Dresselhaus term), and in quantum wells it can
be presented as a sum of linear and cubic contributions.\cite{dk}
Possible asymmetry of a quantum well gives rise to an additional
linear in wavevector contribution to the spin splitting (Rashba
term).\cite{rashba} These two bulk and structural asymmetry terms
have different symmetrical properties and lead to the giant spin
relaxation anisotropy in the heterostructure plane \cite{agw_rev}.
The similar situation is observed in SiGe quantum wells: although
bulk Si and Ge possess inversion center, quantum wells grown from
these materials can lack such a center and allow the spin
splitting of the electronic subbands which can have symmetry of
Dresselhaus or Rashba terms.\cite{golub_i}

One of the main features of D'yakonov-Perel' spin relaxation
mechanism is that it can be suppressed by all processes which
change electron wavevector such as scattering from impurities,
interface microroughness, phonons, electron-electron collisions
\cite{gi_jetp} and because of cyclotron motion of carriers in the
external magnetic field.\cite{ivchenko} Besides, magnetic field
slows down spin relaxation due to Larmor spin precession around
the external field direction.\cite{ivchenko, martin} In Ref.
\onlinecite{janch} it was shown that depending on type of linear
in $\bm k$ terms, the interference between the Larmor and
cyclotron effects will result in either suppression of spin
relaxation due to joint contributions of both effects or to their
mutual compensation leading to partial or full recovery of spin
relaxation rate. However, expressions presented in Ref.
\onlinecite{janch} were obtained in the limit when only one
contribution to the spin splitting is present, moreover as we will
see below they predict incorrect dependence of the transverse
relaxation time $T_2$ on magnetic field orientation.

The aim of present paper is to develop a consistent theory of
D'yakonov-Perel' spin relaxation in classical magnetic fields in
presence of both Dresselhaus and Rashba splittings. In contrast to
Ref. \onlinecite{janch} the expressions for all components of spin
relaxation tensor will be obtained. When both Dresselhaus and
Rashba terms are comparable in magnitude the spin relaxation in
the quantum well plane is strongly anisotropic and correct
definitions of longitudinal $T_1$ and transverse $T_2$ spin
relaxation times is required. We present an expression for spin
susceptibility with allowance for the relaxation anisotropy and
show how electron spin resonance (ESR) linewidth is connected with
spin relaxation tensor components. We analyse recent experiments
on ESR in SiGe quantum wells\cite{janch} and extract electron spin
splitting from the angular dependence of the linewidth.

\section{Theory}
Spin-dependent contribution $\mathcal H_{SO}$ in the effective
Hamiltonian consists of two components. First, in the asymmetrical
quantum wells there exists so-called Rashba term \cite{rashba}
\begin{equation}\label{rashba}
\mathcal H_{R} = \alpha (\sigma_x k_y - \sigma_y k_x),
\end{equation}
where constant $\alpha$ is nonzero due to the heteropotential
asymmetry, $\sigma_i$ ($i=x,y,z$) are Pauli matrices. Axes $x$ and
$y$ are directed along $[100]$ and $[010]$ respectively.

Second contribution has a Dresselhaus form\cite{dk}
\begin{equation}\label{dress}
\mathcal H_{D} = \beta_1 (\sigma_y k_y - \sigma_x k_x) + \beta_3
(\sigma_x k_x k_y^2 - \sigma_y k_y k_x^2).
\end{equation}
In the quantum wells grown from zinc-blende lattice materials
$\beta_3$ is the bulk spin-orbit
constant.\cite{dk,ivch_book,agw_rev} Even if $\beta_3=0$ (like in
SiGe quantum wells) $\beta_1$ is nonzero due to the asymmetry of
chemical bonds at heterointerfaces.\cite{golub_i,rk}

Sum of Rashba (\ref{rashba}) and Dresselhaus-like (\ref{dress})
terms can be conveniently represented as
\begin{equation}\label{so}
\mathcal H_{SO} = \mathcal H_{D} + \mathcal H_{R} =
\frac{\hbar}{2} \bm \sigma \cdot \bm \Omega_{\bm k},
\end{equation}
where components of vector $\bm \sigma$ are Pauli matrices and
$\bm \Omega_{\bm k}$ is the effective Larmor frequency of spin
precession in the spin-splitting induced field. One can see that
vector $\bm \Omega_{\bm k}$ contains both first and third
harmonics of the angle between the wavevector $\bm k$ and
$x$-axis. Spin precession around $\bm \Omega_{\bm k}$ results in
spin relaxation while processes changing electron wavevector such
as scattering and cyclotron motion slow it down.

The application of the external magnetic field $\bm B$ leads to
(a) cyclotron motion of carriers in the quantum well plane, its
frequency is given by
\begin{equation}\label{omega_c}
\omega_{\rm C} = \frac{e B_{z}}{mc},
\end{equation}
where $B_{z}$ is normal component of magnetic field, $e$ and $c$
are elementary charge and light velocity, $m$ is the electron
effective mass and (b) Larmor precession of electron spins with
angular velocity $\bm \omega_{\rm L}$. The components of $\bm
\omega_{\rm L}$ can be presented as
\begin{equation}\label{omega_l}
\omega_{{\rm L,}\: i} = \frac{\mu_B}{\hbar} g_{ij} B_j,
\end{equation}
where $\mu_B = e\hbar/2m_0 c$ is Bohr magneton, $m_0$ is free
electron mass and $g_{ij}$ are the components of the $g$-factor
tensor for electrons in the quantum well, and summation over the
repeated indices is omitted. The direction of the vector $\bm
\omega_{\rm L}$ defines the direction of Larmor precession axis.
In what follows it is convenient to introduce vector $\bm
\omega_{\rm C}$ directed along the growth axis with absolute value
being equal to the cyclotron frequency.

We will use density matrix formalism in order to describe spin
relaxation. Electron spin density matrix can be decomposed as
\[
\rho_{\bm k} = f_{\bm k} + {\bm s}_{\bm k} \cdot {\bm \sigma}\:,
\]
where $f_{\bm k}$ is the spin-averaged occupation of the $\bm k$
state and ${\bm s}_{\bm k}$ is the spin vector for electrons in
this state. Kinetic equation for spin distribution ${\bm s}_{\bm
k}$ can be written in the following form \cite{ivchenko,gi_jetp}
\begin{equation}\label{kinetic}
\frac{\partial \bm s_{\bm k}}{\partial t} + \bm s_{\bm k} \times
(\bm \omega_{\rm L} + \bm \Omega_{\bm k} ) +  \hat{ \bm \Lambda}
\bm s_{\bm k} + \bm Q\{ \bm s_{\bm k} \} =0,
\end{equation}
where the second term describes spin precession in the presence of
both external field and spin splitting induced effective magnetic
field Eq. (\ref{so}), the third term comes from wavevector
variations under cyclotron motion with operator $\hat{ \bm
\Lambda}$ defined according to $(\hat{ \bm \Lambda} {\bm s_{\bm
k}})_i  = \bm \omega_{\rm C} [\bm k\times
\partial s_{\bm k, i}/\partial \bm k]$, and the last term is
the collision integral. For simplicity we will consider only
elastic scattering processes and describe them using two times
$\tau_1$ and $\tau_3$, responsible for relaxation of first and
third harmonics of distribution function. \cite{agw_rev} Eq.
(\ref{kinetic}) is valid for classical magnetic fields, quantum
effects on spin relaxation were discussed in Refs.
\onlinecite{burkov, lyub} for strong and weak fields respectively.

Considering spin-splitting as a small perturbation ($\Omega_{\bm
k} \tau_1$, $\Omega_{\bm k} \tau_3 \ll 1$) one can present spin
distribution function $\bm s_{\bm k}$ as
\[
\bm s_{\bm k} = \bm s_{\bm k}^0 + \delta \bm s_{\bm k},
\]
where $\bm s_{\bm k}^0$ is quasi-equilibrium distribution function
and $\delta \bm s_{\bm k}$ is a non-equilibrium correction arising
due to spin splitting. After summation (\ref{kinetic}) over the
wavevectors we arrive to the balance equation describing
relaxation of the total spin $\bm S = \sum_{\bm k} \bm s_{\bm k}$
\begin{equation}\label{balance}
\frac{d \bm S}{dt} + \bm S \times \bm \omega_{\rm L} +\hat{\bm
\Gamma}\bm S =0,
\end{equation}
where the tensor of inverse spin relaxation times $\hat{\bm
\Gamma}$ is defined according to
\begin{equation}\label{gammadef}
\hat{\bm \Gamma}\bm S = \sum_{\bm k} \delta \bm s_{\bm k} \times
\bm \Omega_{\bm k},
\end{equation}
and the non-equilibrium correction $\delta \bm s_{\bm k}$ should
be determined from the solution of Eq. (\ref{kinetic}).

We will be interested in the case of strong enough magnetic field
: $ \Omega_{\bm k}^2 \tau_1$, $ \Omega_{\bm k}^2 \tau_3 \ll
\omega_{\rm L}$. Then all non-diagonal components of the tensor
$\hat{\bm \Gamma}$ associating spin relaxation along external
field and in the transverse plane are zero. Noting that the first
and third harmonics of spin distribution have independent dynamics
we present $\hat{\bm \Gamma}$ in the form
\begin{equation}\label{inverse_sr_tensor}
\hat{\bm \Gamma} = \hat{\bm \Gamma}^{(1)} + \hat{\bm
\Gamma}^{(3)},
\end{equation}
where the upper index denotes the number of angular harmonic.

\begin{figure}[h]
  \includegraphics[width=0.5\textwidth]{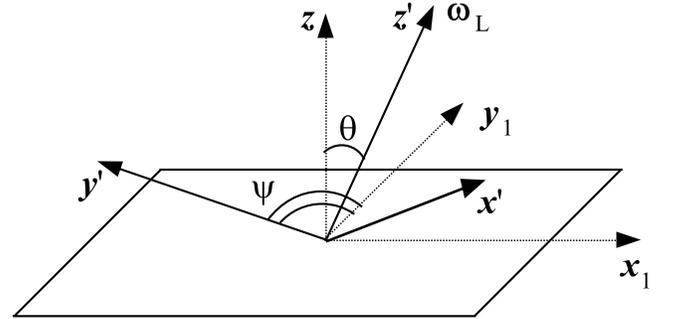}\\
  \caption{Frames of axes. $z \parallel [001]$ is a growth
  axis. Axes $x_1 \parallel [1\bar10]$ and $y_1 \parallel
  [110]$. The frame of axis $x'$, $y'$ and $z'$
  is connected with the external magnetic field: $z'\parallel \bm \omega_{\rm
  L}$, $y'$ is a perpendicular to $\bm \omega_{\rm L}$
  lying in a quantum well plane, and $x'$ is chosen so that $x'$, $y'$ and $z'$
  form a right-hand triple.
  }\label{fig1}
\end{figure}

It is convenient to present components of spin relaxation tensor
in the new frame of coordinates with $z'$ axis directed along
Larmor frequency vector $\bm \omega_{\rm L}$, and $y'$ axis lying
in the quantum well plane and being perpendicular to the $\bm
\omega_{\rm L}$ (Fig. \ref{fig1}). Let $\theta$ be the angle
between $z'$ and $z$ and $\psi$ be the angle between $y'$ and
$y_1\parallel [110]$. One can show that for electrons with fixed
energy $E$ the non-zero components of tensors $\hat{\bm
\Gamma}^{(1)}$ и $\hat{\bm \Gamma}^{(3)}$ can be written as
\begin{widetext}
\begin{equation}\label{gammas}
\Gamma^{(1)}_{z'z'} = \frac{2 \tau_1 k^2}{\hbar^2}
\frac{N_{1,z'z'}} {D_{1,+} D_{1,-}}\:, \quad \Gamma^{(1)}_{ii} =
\frac{2 \tau_1 k^2}{\hbar^2} \frac{N_{1,ii}} {1 + \omega_{\rm C}^2
\tau_1^2}\:, \quad \Gamma^{(1)}_{x'y'} = -\frac{4\tau_1 k^2
\alpha\beta}{\hbar^2} \frac{\cos{\theta}\sin{2\psi}}{1 +
\omega_{\rm C}^2 \tau_1^2}\:,
\end{equation}

\[
\Gamma^{(3)}_{z'z'} = \frac{2\tau_3 k^6 \beta_3^2}{8 \hbar^2}
\frac{N_{3,z'z'}} {D_{3,-}}\:, \quad \Gamma^{(3)}_{x'x'} =
\frac{\tau_3 k^6 \beta_3^2}{8 \hbar^2} \left\{ \frac{1}{1 +
9\omega_{\rm C}^2 \tau_3^2} + \frac{\sin^2 {\theta}[1 +
(\omega_{\rm L}^2 + 9\omega_{\rm C}^2)\tau_3^2]}{D_{3,+}D_{3,-}}
\right\}\:,
\]

\[
\Gamma^{(3)}_{y'y'} = \frac{\tau_3 k^6 \beta_3^2}{8 \hbar^2(1 +
9\omega_{\rm C}^2 \tau_3^2)} \left\{ 1 - \frac{\sin^2 {\theta}[1 +
(\omega_{\rm L}^2 - 27\omega_{\rm C}^2)\tau_3^2] \omega_{\rm
L}^2\tau_3^2 }{D_{3,+}D_{3,-}} \right\}\:.
\]
Here $k=\sqrt{2mE}/\hbar$, $i=x',y'$, $\beta=\beta_1 -\beta_3
k^2/4$ is a coefficient at first harmonics in Dresselhaus
splitting (\ref{dress}), $D_{j,\pm} = 1 + (\omega_{\rm L} \pm
j\omega_{\rm C})^2\tau_j^2$, where $j=1,3$, and
\begin{equation}\label{ns}
N_{1,z'z'} = \alpha^2 \left[1+ \cos^2{\theta} + (\omega_{\rm C} +
\omega_{\rm L} \cos{\theta})^2 \tau_1^2 + (\omega_{\rm
C}\cos{\theta} + \omega_{\rm L})^2 \tau_1^2 \right]
\end{equation}
\[
+\beta^2 \left[1+ \cos^2{\theta} + (\omega_{\rm C} - \omega_{\rm
L} \cos{\theta})^2 \tau_1^2 + (\omega_{\rm C}\cos{\theta} -
\omega_{\rm L})^2 \tau_1^2 \right]+ 2\alpha\beta
\cos{2\psi}\sin^2{\theta} [1+(\omega_{\rm C}^2 + \omega_{\rm L}^2)
\tau_1^2]\:,
\]
\[
N_{1,x'x'} = (\alpha^2 + \beta^2) \left\{ 1 + \sin^2{\theta} (1
+\omega_{\rm C}^2 \tau_1^2) [1+(\omega_{\rm C}^2 + \omega_{\rm
L}^2) \tau_1^2)](D_{1,+}D_{1,-})^{-1} \right\}+ 2\alpha\beta
\cos{2\psi} \:,
\]
\[
N_{1,y'y'} = (\alpha^2 + \beta^2) \left\{ 1 - \sin^2{\theta}
[{1+(\omega_{\rm L}^2 - 3\omega_{\rm C}^2)
\tau_1^2}](D_{1,+}D_{1,-})^{-1} \omega_{\rm L}^2 \tau_1^2
\right\}- 2\alpha\beta \cos{2\psi} \:,
\]
\[
N_{3,z'z'} = 1 - \sin^2{\frac{\theta}{2}}\left\{1 + \left
[{\cos{\theta} \biggl( 1 + (\omega_{\rm L}^2 + 9\omega_{\rm
C}^2)\tau_3^2 \biggr) - 6 \omega_{\rm L} \omega_{\rm C} \tau_3^2 }
\right]D_{3,+}^{-1} \right\}\:.
\]
Eqs. (\ref{gammas}) and (\ref{ns}) describe spin relaxation of
two-dimensional electron gas subject to classical magnetic field.
These formulae are valid for arbitrary ratio of Dresselhaus and
Rashba contributions and for arbitrary orientation of the external
field. Generalization of Eqs. (\ref{gammas}), (\ref{ns}) for the
case of any given energy distribution of electrons can be carried
out in the standard way.\cite{agw_rev}

\section{Results and discussion}
\subsection{Longitudinal relaxation}
We will start the analysis of the expressions Eqs. (\ref{gammas})
and (\ref{ns}) for the spin relaxation tensor components in the
case when magnetic field is normal to the quantum well structure,
i.e. $\theta=0$. Then the expression for the longitudinal spin
relaxation rate $1/T_1 = \Gamma_{zz} = \Gamma^{(1)}_{zz} +
\Gamma^{(3)}_{zz}$ reduces to
\begin{equation}\label{normal}
\frac{1}{T_1}= \frac{4\tau_1 k^2}{\hbar^2}
\left[\frac{\alpha^2}{1+(\omega_{\rm L} - \omega_{\rm
C})^2\tau_1^2} + \frac{\beta^2}{1+(\omega_{\rm L} + \omega_{\rm
C})^2\tau_1^2} \right]
 + \frac{\tau_3 k^6 \beta_3^2}{4\hbar^2}
\frac{1}{1+(\omega_{\rm L} - 3\omega_{\rm C})^2\tau_3^2}\:.
\end{equation}
\end{widetext}

Eq. (\ref{normal}) shows that external magnetic field has
different effects on spin relaxation depending on the dominant
type of spin splitting. When Rashba term is dominant ($\beta_3=0$,
$\beta=0$, $\alpha\ne0$) and electron $g$-factor is positive the
Larmor and cyclotron effects partially compensate each other. In
the case when first harmonics in Dresselhaus splitting is dominant
($\beta\gg \beta_3 k^2$, $\alpha=0$) both Larmor and cyclotron
effects jointly slow down spin relaxation. If the spin splitting
is dominated by the third harmonics then Larmor and cyclotron
effects compensate each other.

\begin{figure}[t]
  \centering
  \includegraphics[width=0.9\linewidth]{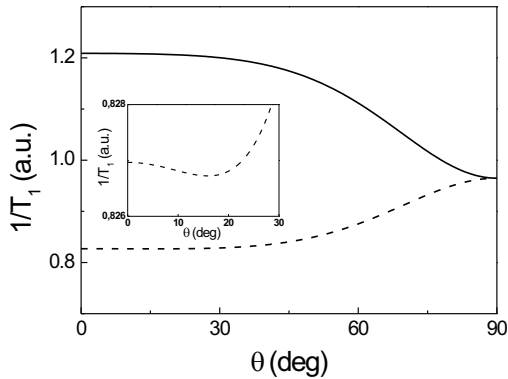}\\
  \caption{Longitudinal spin relaxation rate vs. magnetic field
  tilt angle $\theta$ for SiGe based quantum well. Spin relaxation rate is measured in the
  units $2\tau_1 k^2\delta^2/\hbar^2$ where $\delta$ is a
  coefficient at the first harmonics of spin splitting. Solid curve
  corresponds to Rashba term ($\alpha=\delta$, $\beta=0$),
  dashed curve to Rashba term ($\alpha=0$, $\beta=\delta$). The
  contribution of the third harmonics is neglected. At $\theta=0$
  the parameter $\omega_{\rm C} \tau_1=1$. The parameters of calculation were taken
  from Ref. \onlinecite{ivch_book}: $g=2$, $m=0.196m_0$. Inset shows nonmonotonous behavior
  of $1/T_1(\theta)$ when Dresslhaus term dominates.
  }\label{fig2}
\end{figure}

Such interference behavior is a result of the fact that according
to Eq. (\ref{gammadef}) spin relaxation rate is determined by the
vector product of nonequilibrium correction $\delta \bm s_{\bm k}$
and effective spin precession frequency $\bm \Omega_{\bm k}$. If
no magnetic field is applied then $\delta \bm s_{\bm k}$ is
orthogonal to $\bm \Omega_{\bm k}$ and spin relaxation rate is
fastest. External magnetic field results in the precession of
in-plane spin component $\delta \bm s_{\bm k}$ around Larmor
vector ${\bm \omega}_{\rm L}$ and in cyclotron rotation of the
$\bm \Omega_{\bm k}$. If these rotations are not synchronous then
vectors $\delta \bm s_{\bm k}$ and $\bm \Omega_{\bm k}$ are no
longer orthogonal and spin relaxation rate becomes slower. In the
case of synchronous rotation of in-plane spin and effective
magnetic field the spin relaxation is not suppressed. It is
important to note that the direction of cyclotron rotation of $\bm
\Omega_{\bm k}$ depends on the dominant spin splitting mechanism.
If Rashba term is largest then effective magnetic field $\bm
\Omega_{\bm k}$ is perpendicular to the wavevector and rotates in
the same direction as $\bm k$. Thus, spin relaxation becomes
slower by the factor of $1+(\omega_{\rm L} - \omega_{\rm
C})^2\tau_1^2$ and Larmor and cyclotron effects partially
compensate each other provided electron $g$-factor being positive.
If spin splitting is due to the first harmonics of Dresselhaus
term then cyclotron rotation of $\bm \Omega_{\bm k}$ and Larmor
precession at positive $g$-factor have opposite directions and
slow down spin relaxation by the factor of $1+(\omega_{\rm L} +
\omega_{\rm C})^2\tau_1^2$. When third harmonics dominates, the
rotation of $\bm \Omega_{\bm k}$ and Larmor precession take place
in the same direction but the rotation frequency of $\bm
\Omega_{\bm k}$ is three times higher than cyclotron one.

Figure \ref{fig2} shows longitudinal spin relaxation rate $1/T_1 =
\Gamma_{z'z'}$ dependence on the magnetic field tilt angle
$\theta$ calculated for SiGe structure. The case of dominant
Rashba term is shown by a solid line, and dashed line presents the
results when spin splitting is determined by Dresselhaus term. The
contribution of the third harmonics is neglected in all
calculations. The qualitatively different angular dependencies of
$1/T_1$ are clearly seen. When Rashba contribution is a main one
the spin relaxation rate decreases with increase of in-plane
component of magnetic field because tilting of the field results
in the decrease of the cyclotron frequency $\omega_{\rm C}$. When
Dresselhaus term is dominant the dependence $1/T_1(\theta)$ is
nonmonotonous (see inset): first, spin relaxation slows down owing
to the decrease of projection of $\bm \Omega_{\bm k}$ on the plane
perpendicular to the magnetic field. At larger $\theta$ spin
relaxation accelerates due to cyclotron frequency decrease.

\begin{figure}[t]
  \includegraphics[width=0.95\linewidth]{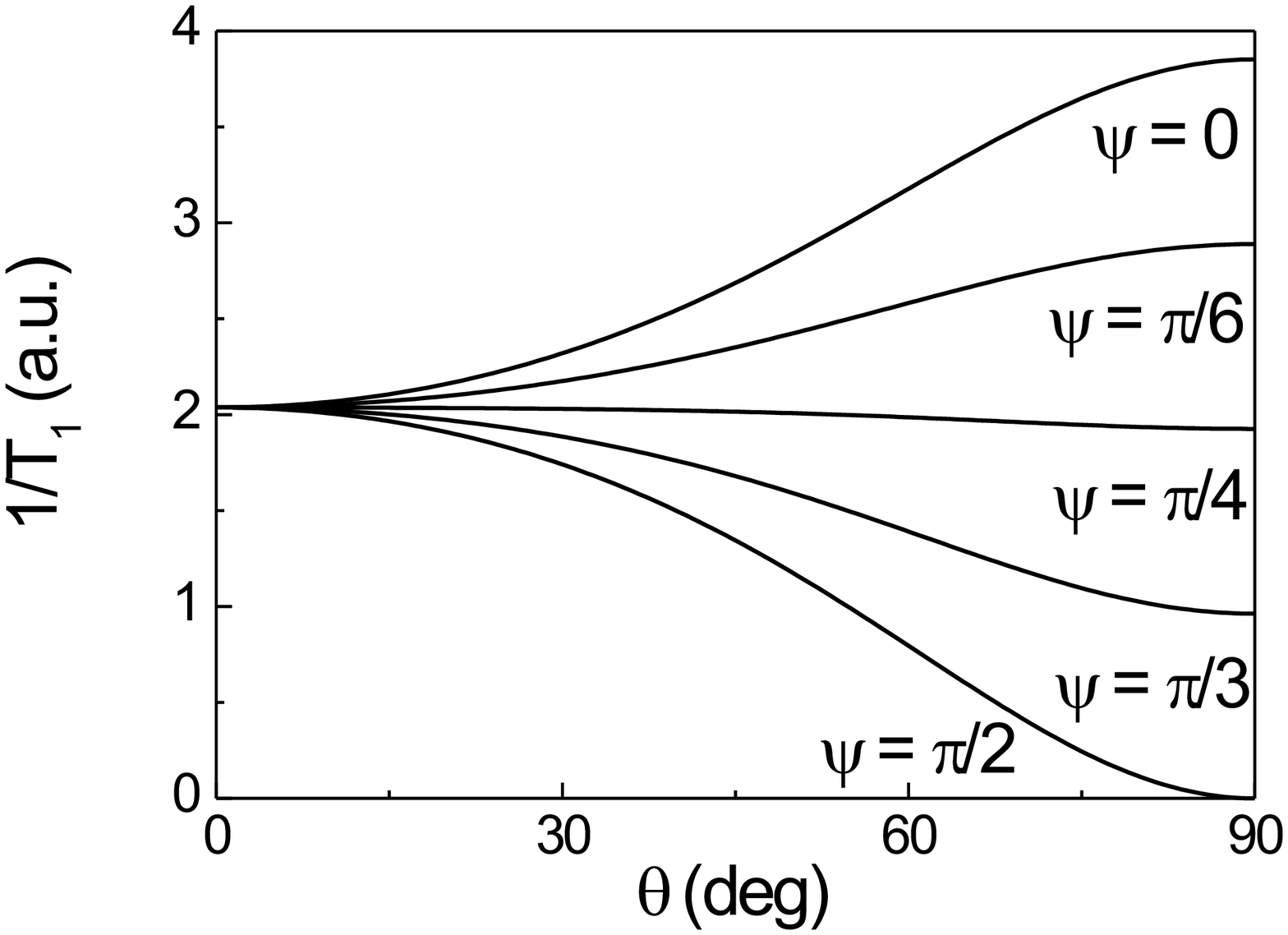}\\
  \caption{Longitudinal spin relaxation rate vs. magnetic field
  tilt angle $\theta$ in the case of equal Dresselhaus and Rashba
  terms($\alpha=\beta$) for different orientations of magnetic field.
  Spin relaxation rate is presented in units $2\tau_1 k^2\alpha^2/\hbar^2$.
  The values of parameters used in the
  calculation are the same as for Fig.~\ref{fig2}.
  }\label{fig3}
\end{figure}

Figure \ref{fig3} shows longitudinal spin relaxation rate for
different orientations of magnetic field in the case when Rashba
and Dresselhaus terms have equal strengths ($\alpha=\beta$).
Different curves correspond to different angles between magnetic
field and main axes of the structure. One can see that spin
relaxation rate is highest at $\psi=0$, i.e. for magnetic field
inclined to the $x_1\parallel [1\bar1 0]$ axis and is lowest at
$\psi=\pi/2$ (magnetic field inclined to $y_1
\parallel [1 1 0]$). This result is a direct consequence of the
interference of linear in $\bm k$ spin dependent terms resulting
in the suppression at $\alpha=\beta$ of the spin splitting along
$[1\bar1 0]$.\cite{agw_rev} For the same reason spin relaxation
rate calculated with allowance for the first harmonics of the
wavevector only goes to zero when magnetic field is directed along
$[1 1 0]$, the account for the third harmonics in spin splitting
leads to the finite spin lifetime along $y_1$. \cite{agw_rev}

\begin{figure*}[t]
  \centering
  \includegraphics[width=0.9\textwidth]{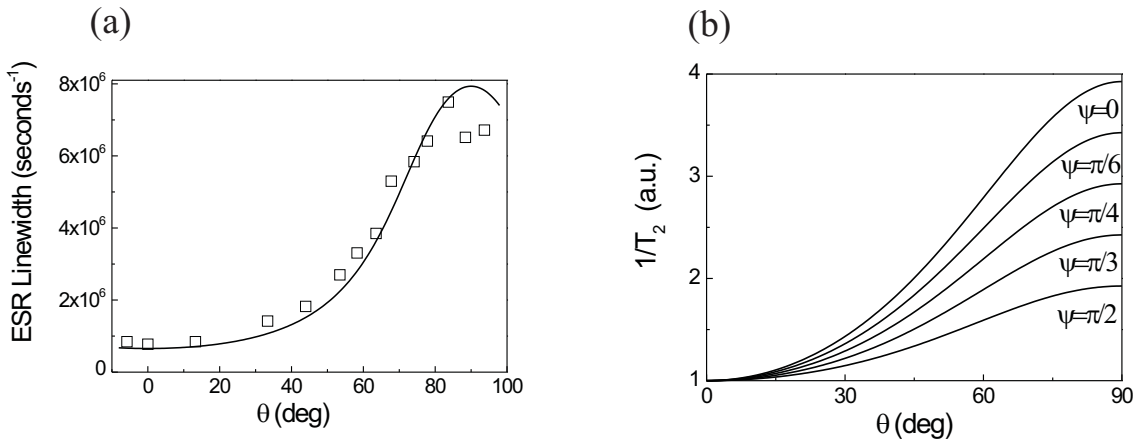}\\
  \caption{Transverse relaxation time vs. magnetic field tilt angle $\theta$
  for SiGe structure. (a) Open squares show ESR linewidth measured in Ref. \onlinecite{janch}.
  Solid curve presents the results of calculation when Rashba term
  dominates. Parameters $\omega_{\rm C} = 2.96 \times
  10^{11}$~s$^{-1}$ at normal magnetic field orientation,
  $\omega_{\rm L} = 5.93 \times 10^{10}$~s$^{-1}$,
  and momentum relaxation time $\tau_1 =
  10^{-11}$~s were taken from experiment. Spin precession frequency is
  $\Omega_{k_F} = \alpha k_F/\hbar = 4 \times
  10^8$~s$^{-1}$. (b) Dresselhaus and Rashba terms are equal ($\alpha=\beta$).
  Spin relaxation rate is measured in the units of $\tau_1 k^2\alpha^2/\hbar^2$.
  Parameters of calculation are the same as for Fig. \ref{fig2}a.
  }\label{fig4}
\end{figure*}

Different behavior of spin relaxation rate for different values of
$\psi$ can be explained taking into account that in normal
magnetic field spin relaxation is suppressed mostly by cyclotron
effect since $\omega_{\rm L} / \omega_{\rm C} = g m/2m_0 \approx
0.2$. With increase of $\theta$ cyclotron frequency decreases, and
if $-\pi/4 \le \psi \le \pi/4$ effective spin splitting increases.
Therefore spin relaxation becomes faster if $-\pi/4 \le \psi <
\pi/4$. At $\psi=\pi/4$ and $\omega_{\rm C} \tau_1 = 1$ spin
relaxation rate should not depend of tilt angle if one completely
neglects Larmor effect. With tilting of magnetic field cyclotron
frequency decreases and Larmor effect becomes more important
resulting in slight decrease of relevant curve ($\psi=\pi/4$). If
$\pi/4 < \psi \le 3\pi/4$ then competition between decrease of
$\omega_{\rm C}$, decrease of effective spin splitting and Larmor
effect is possible leading to nonmonotonous angular dependence of
the spin relaxation rate. However at $\omega_{\rm C} \tau_1 = 1$
taken for calculation presented in Fig.\ref{fig3}b spin relaxation
slows down.

\subsection{Transverse relaxation}

Now we turn to the analysis of spin relaxation in the plane
perpendicular to the external field. In the ESR experiments the
spin susceptibility is measured which in the limit of low
microwave power has a following form
\begin{equation}\label{chi}
\chi \propto \frac{1}{\omega_{\rm L} - \omega + {\rm i} /T_2 }
\end{equation}
where $\omega$ is microwave frequency and $T_2$ is a transverse
relaxation time. Eq. (\ref{chi}) is valid when microwave frequency
is close to Larmor one, small renormalization of resonance
frequency $\sim T_2^{-1}$ is neglected. Transverse relaxation time
can be expressed through the components of spin relaxation tensor
as
\begin{equation}\label{T2}
1/T_2 = \frac{\Gamma_{x'x'} + \Gamma_{y'y'}}{2},
\end{equation}
and ESR half-maximum full width $\Delta \omega = 2/T_2$. Eq.
(\ref{T2}) generalizes results of Refs. \onlinecite{janch,tahan}
for the case of arbitrary orientation of magnetic field and
anisotropic spin relaxation. One can see that the combination
$\Gamma_{x'x'} + \Gamma_{y'y'}$ is invariant to the choice of axes
in the transverse plane, therefore resonant peak width is
independent of microwave field polarization.

If the external field is oriented along the quantum well growth
axis then $1/T_2$ has the form
\begin{equation}\label{eqt2}
1/T_2 = \frac{2\tau_1 k^2}{\hbar^2} \frac{\alpha^2 + \beta^2}{1 +
\omega_{\rm C}^2 \tau_1^2} + \frac{\tau_3 k^6}{8\hbar^2}
\frac{\gamma^2}{1 + 9\omega_{\rm C}^2 \tau_1^2}.
\end{equation}
In the particular case of $\beta=0$ and $\omega_{\rm C}=0$ this
expression coincides with equations obtained in Ref.
\onlinecite{tahan}. However it is in sharp disagreement with
results of Ref. \onlinecite{janch}, where zero spin relaxation
rate is predicted for normal magnetic field. Those results were
based on the assumption that transverse spin relaxation is
governed by the normal component of $\bm \Omega_{\bm k}$. This
assumption is invalid, since spin relaxation along, say, $x'$ is
governed by both $\Omega_{\bm k,y'}$ and $\Omega_{\bm k,z'}$.

Figure \ref{fig4} shows the results of calculation of the
transverse spin relaxation rate $1/T_2$ in SiGe structure versus
external magnetic field orientation. The transverse relaxation
rate monotonously increase with tilt angle (solid curve on Fig.
\ref{fig4}a). It comes from the fact that spin relaxation in the
plane normal to magnetic field is suppressed owing to cyclotron
effect, see Eqs. (\ref{gammas}) and (\ref{eqt2}). Cyclotron
frequency is maximal if magnetic field is normal to the quantum
well plane and decreases proportionally to $\cos{\theta}$ with
tilting magnetic field.

Open squares on Fig. \ref{fig4}a show the ESR linewidth measured
in SiGe quantum well structure.\cite{janch} The parameters of
calculation were taken equal to the experimental ones:
$\omega_{\rm C} = 2.96 \times 10^{11}$~s$^{-1}$ at normal magnetic
field, $\omega_{\rm L} = 5.93 \times 10^{10}$~s$^{-1}$, the
momentum relaxation time $\tau_1 = 10^{-11}$~s. We used the
effective frequency of spin precession on the Fermi level
$\Omega_{k_F} = \alpha k_F/\hbar$ as a fitting parameter. The best
agreement is at $\Omega_{k_F} = 4\times 10^8$~s$^{-1}$. The order
of magnitude of $\Omega_{k_F}$ agrees with estimations of Ref.
\onlinecite{tahan}. Obtained splitting is lower than value
$6.3\times 10^8$~s$^{-1}$ presented in Ref. \onlinecite{janch},
because fitting in that paper was done with incorrect dependence
$1/T_2 (\theta)$.

In the case when Dresselhaus and Rashba terms are equal in
magnitude ($\alpha=\beta$), the dependence of transverse
relaxation rate $1/T_2(\theta)$ on magnetic field tilt angle is
qualitatively the same. The slope depends on the in-plane rotation
angle $\psi$ reflecting anisotropy of the system.

In conclusion, we have studied in detail effects of magnetic field
on electron spin relaxation in SiGe quantum wells. We have shown
that, depending on electron $g$-factor sign and dominant spin
splitting mechanism, Larmor and cyclotron effects are either
compete with each other or jointly suppress spin relaxation. The
expression for ESR linewidth was derived for the case of
anisotropic transverse relaxation. Fitting by the derived
expressions the experimental data we obtained spin splitting value
for SiGe structure studied in Ref. \onlinecite{janch}.

\begin{acknowledgements}
Author is grateful to L.E.~Golub and E.L.~Ivchenko for valuable
discussions. Work was partially supported by RFBR and scholarship
of ``Dynasty'' foundation -- ICFPM.
\end{acknowledgements}

\newpage

\end{document}